\documentclass[twocolumn,showpacs,preprintnumbers,amsmath,amssymb,nofootinbib]{revtex4}

\usepackage{overpic}
\usepackage{graphicx}
\usepackage{epsfig}
\usepackage{dcolumn}
\usepackage{bm}

\newcommand{\eff}{\varepsilon}

\newcommand{\etap}{\eta^{\prime}}

\newcommand{\be}{\begin{enumerate}}
\newcommand{\ee}{\end{enumerate}}
\newcommand{\bi}{\begin{itemize}}
\newcommand{\ei}{\end{itemize}}

\newcommand{\jpsi}{J/\psi}

\newcommand{\pip}{\pi^+}
\newcommand{\pin}{\pi^-}
\newcommand{\pio}{\pi^0}
\newcommand{\g}{\gamma}
\newcommand{\ar}{\rightarrow}

\def\Journal#1&#2&#3(#4){#1{\bf #2}, #3 (#4)}

\def\bec{\begin{center}}
\def\eec{\end{center}}

\begin{document}
\title{ \bf \boldmath Measurements 
of  $\jpsi$ decays into $\omega\pio$, $\omega\eta$, and $\omega\etap$ }

\author{
M.~Ablikim$^{1}$,              J.~Z.~Bai$^{1}$,               Y.~Ban$^{11}$,
J.~G.~Bian$^{1}$,              X.~Cai$^{1}$,
H.~F.~Chen$^{16}$,
H.~S.~Chen$^{1}$,              H.~X.~Chen$^{1}$,
J.~C.~Chen$^{1}$,
Jin~Chen$^{1}$
% ...
\\(BES Collaboration)\\
\vspace{0.2cm}
$^{1}$ Institute of High Energy Physics, Beijing 100049, People's Republic
of China\\
% ...
$^{11}$ Peking University, Beijing 100871, People's Republic of China\\
% ...
$^{16}$ University of Science and Technology of China, Hefei 230026,
People's Republic of China\\
% ...
}

\noindent\vskip 0.2cm 
\begin{abstract}
Based on $5.8 \times 10^7 \jpsi$ events collected with BESII at the
Beijing Electron-Positron Collider (BEPC), the decay branching
fractions of $\jpsi\to\omega\pio$, $\omega\eta$, and $\omega\etap$ are
measured using different $\eta$ and $\etap$ decay modes. The results are
higher than previous measurements.  The $\omega\pio$
electromagnetic form factor is also obtained.

\end{abstract}

\maketitle 

\section{Introduction}   \label{introd} 

The decay of the $J/\psi$ into a vector and pseudoscalar (VP) meson
pair proceeds via $c \bar{c}$ annihilation to three gluons in strong
decays and to one virtual photon in electromagnetic decays.  A full
set of $J/\psi \to V P$ measurements allows one to systematically
study the quark-gluon content of pseudoscalar mesons and SU(3)
breaking, as well as determine the contribution of different amplitudes
to the decay rates in two-body $J/\psi$ decays~\cite{theory}.
Measurements of purely electromagnetic $\jpsi$ decays can be used to
calculate the electromagnetic form factors involved; those form
factors are used to test QCD inspired models of the mesonic wave
function.  MARKIII~\cite{mark2,mark3} and DM2~\cite{dm2} measured many
$J/\psi \to V P$ branching fractions and obtained the $\eta-\eta'$
mixing angle, the quark content of the $\eta$ and $\eta'$, and much
more.

Recently, a sample of $5.8 \times 10^7 \jpsi$ events, which offers a
unique opportunity to measure the full set of $J/\psi \to V P$ decays
precisely, was accumulated with the upgraded Beijing Spectrometer
(BESII)~\cite{besii}.  In an earlier analysis based on this data set
and using a GEANT3 based Monte-Carlo with a careful simulation of
detector response, the branching fraction of $\jpsi\ar\pip\pin\pio$
was measured to be $(2.10\pm0.12)\%$~\cite{rhopi2}, which is higher
than the PDG~\cite{pdg2004} value by about 30\%.  This indicates a
higher branching fraction for $\jpsi\ar\rho\pi$ than those from older
experiments~\cite{pdg2004}. Furthermore, the branching ratios of
$J/\psi \to \phi P(\pio,\eta,\etap)$~\cite{phip} are also higher than
old experimental results. Therefore, remeasuring the branching
fractions of all $\jpsi\ar VP$ decay modes becomes very important.  In
this paper, $J/\psi \to \omega \pi^0$, $\omega \eta$, and $\omega
\etap$ are studied using different $\eta$ and $\etap$ decay modes.
%based on the BESII $5.8 \times 10^7 \jpsi$ events.

\section{The BES Detector}  \label{BESD} 

The upgraded Beijing Spectrometer detector (BESII) is located at the
Beijing Electron-Positron Collider (BEPC). BESII is a large
solid-angle magnetic spectrometer which is described in detail in
Ref.~\cite{besii}.  The momentum of  charged particles is determined
by a 40-layer cylindrical main drift chamber (MDC) which has a
momentum resolution of $\sigma_{p}$/p=$1.78\%\sqrt{1+p^2}$ ($p$ in
GeV/c).  Particle identification is accomplished using specific
ionization ($dE/dx$) measurements in the drift chamber and
time-of-flight (TOF) information in a barrel-like array of 48
scintillation counters. The $dE/dx$ resolution is
$\sigma_{dE/dx}\simeq8.0\%$; the TOF resolution for Bhabha events is
$\sigma_{TOF}= 180$ ps.  Radially outside of the time-of-flight
counters is a 12-radiation-length barrel shower counter (BSC)
comprised of gas tubes interleaved with lead sheets. The
BSC measures the energy and direction of photons with resolutions of
$\sigma_{E}/E\simeq21\%\sqrt{E}$ ($E$ in GeV), $\sigma_{\phi}=7.9$
mrad, and $\sigma_{z}=2.3$ cm. The iron flux return of the magnet is
instrumented with three double layers of proportional counters
that are used to identify muons.

A GEANT3 based Monte Carlo (MC) package (SIMBES)~\cite{pid} with
detailed consideration of the detector performance is used.  The
consistency between data and MC has been carefully checked in many
high purity physics channels, and the agreement is reasonable.  More
details on this comparison can be found in
Ref.~\cite{pid}. In this analysis, the detection efficiency and mass
resolution for each decay mode are obtained from a MC simulation,
which takes into account the angular distributions appropriate for the
different final states~\cite{rhopi}.

\section{analysis}
\label{analysis}
%\section{Event Selection}
In this analysis, the $\omega$ meson is observed in its $\pip\pin\pio$ decay
mode, and the pseudoscalar mesons are detected in the modes:
$\pio\ar\g\g$; $\eta\ar\g\g$, $\g\pip\pin$, and $\pip\pin\pio$; and
$\etap\ar\g\pip\pin (\g\rho)$ and
$\pip\pin\eta~(\eta\ar\g\g)$.  Using the different $\eta$ and $\etap$ decay
modes allows us to cross check our measurements, as
well as obtain higher precision. Possible final states of
$\jpsi\ar \omega\pio,~\omega\eta$, and $\omega\etap$ are then
$\pip\pin\g\g\g\g$,
$\pip\pin\pip\pin\g\g\g$, and $\pip\pin\pip\pin\g\g\g\g$.

Candidate events are
required to satisfy the following common selection criteria:

\begin{enumerate}
\item Events must have two or four good charged tracks with zero
net charge.  A good charged track is a
track that is well fitted to a helix, originates from the interaction
region of R$_{xy}< $0.02 m and $|z| <$ 0.2 m, and has a polar angle,
$\theta$, in the range $|\cos \theta| <$ 0.8.  R$_{xy}$ is the
distance from the beamline to the point of closest approach of the
track to the beamline, and $|z| $ is the distance along the beamline to
this point from the interaction point.

\item Candidate events should have at least the minimum number of
isolated photons associated with the different final states, unless
otherwise specified.  Isolated photons are those that have energy
deposited in the BSC greater than 60 MeV, the angle between the photon
entering the BSC and the shower development direction in the BSC less
than 30$^\circ$, and the angle between the photon and any charged
track larger than 10$^\circ$.

\item  For each charged track in an event, $\chi^{2}_{PID}(i)$ is
  determined using
both $dE/dx$ and TOF information:
\begin{center}
 $\chi^{2}_{PID}(i)$=$\chi^{2}_{dE/dx}(i)$+$\chi^{2}_{TOF}(i),$
\end{center}
where $i$ corresponds to the particle hypothesis.  A charged track is
identified as a $\pi$ if $\chi^{2}_{PID}$ for the $\pi$ hypothesis is
less than those for the $K$ or $p$ hypotheses.  For the channels
studied, at least one charged track must be identified as a pion.

\item The selected events are subjected to four constraint (4-C)
  kinematic fits, unless otherwise specified.  When there are more
  than the minimum number of photons in an event, all combinations are
  tried, and the combination with the smallest $\chi^{2}$ is
  retained. The $\chi^2$ values required in the selection of events
  below are based on the optimization of $S/\sqrt{S+B}$, where the $S$
  and $B$ are the expected signal and background, respectively.

\end{enumerate}

The branching fraction is calculated using

\begin{eqnarray*}
\lefteqn{B(\jpsi\ar \omega P) = }  \\ 
&&\frac{N_{obs}}{N_{\jpsi}\cdot
\eff\cdot B(\omega\ar \pip\pin\pio)\cdot B(P\ar X)\cdot B(\pio\ar\g\g)}
\label{forbr}
\end{eqnarray*}
where $N_{obs}$ is the number of events observed, $N_{\jpsi}$ is the
total number of $\jpsi$ events, $(5.77\pm 0.27)\times
10^7$~\cite{fangss}, $\eff$ is the detection efficiency obtained from
MC simulation which takes into account the angular
distributions~\cite{rhopi}, and $B(\omega\ar\pip\pin\pio)$ and $B(P\ar
X)$ are the branching fractions, taken from the Particle Data Group
(PDG)~\cite{pdg2004}, of $\omega\ar\pip\pin\pio$ and the pseudoscalar $P$ to 
$X$ final states, respectively.

\subsection{\boldmath $\jpsi\ar\omega\g\g$}
\begin{figure}[htbp]
\centerline{\hbox{\psfig{file=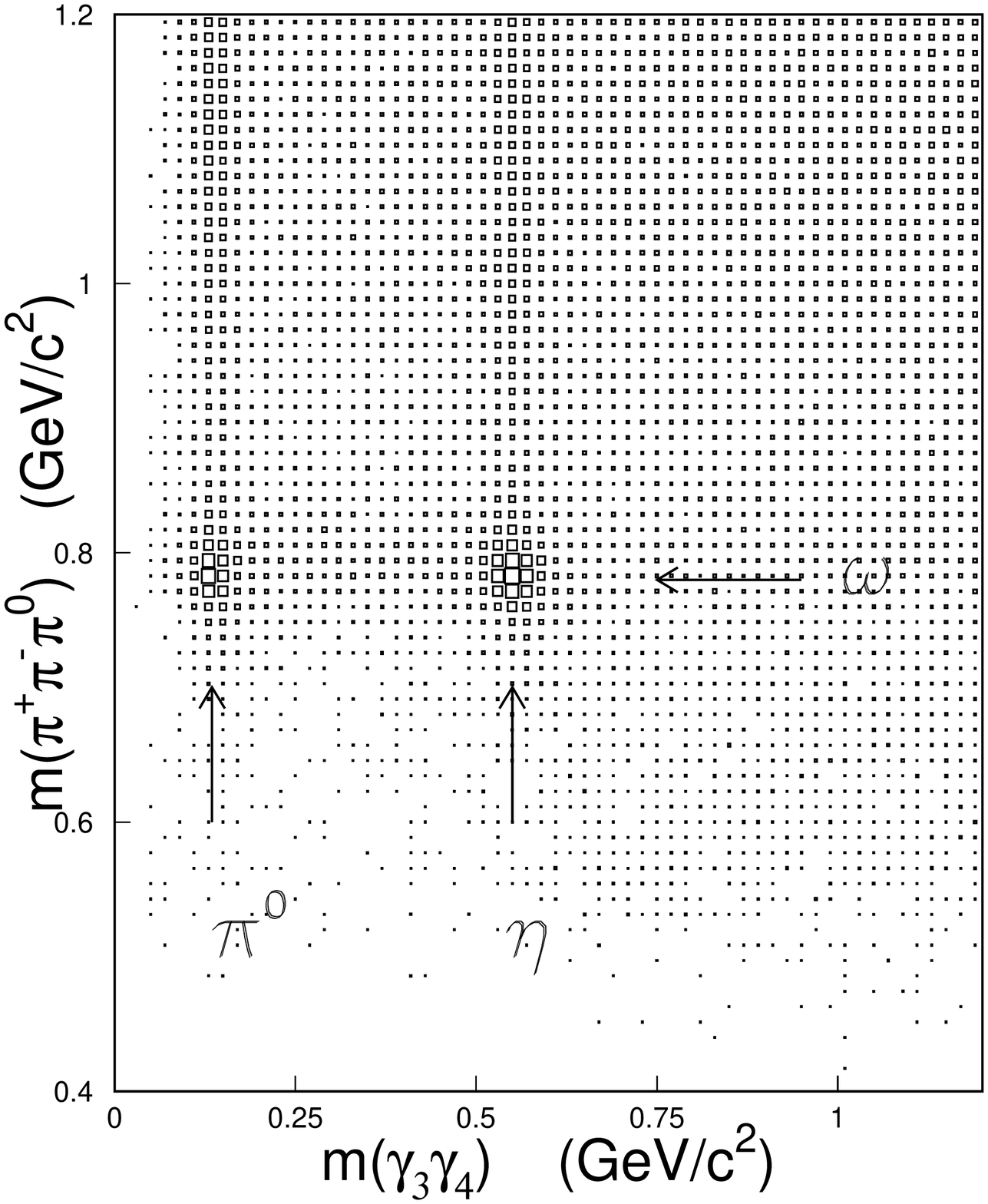,width=5.5cm,height=5.5cm}}
} 
\caption{Scatter plot of $m_{\g_3\g_4}$ versus $m_{\pip\pin\pi^0}$ for 
$\jpsi\ar\pip\pin\g\g\g\g$ candidate events.}
\label{domegapi0}
\end{figure}

Events with two oppositely charged tracks and at least four isolated
photons are selected.  A 4C-fit is performed to the $\pip \pin \gamma
\gamma \gamma \gamma$ hypothesis, and $\chi^2 < 15$ is required.
There are six $\gamma \gamma$ combinations to test for consistency
with the $\pi^0$ mass.  Looping over all combinations, we calculate
$m_{\pip\pin\g_1\g_2}$ for combinations satisfying
$|m_{\g_1\g_2}-0.135|<0.04$ GeV/c$^2$, denoted as $m_{\pip\pin\pi^0}$,
and plot $m_{\pip\pin\pi^0}$ versus $m_{\g_3\g_4}$ in
Fig.~\ref{domegapi0}, where clear $\pio$ and $\eta$ signals are seen.

\subsubsection{$\jpsi\ar\omega\pio$}

\begin{figure}
\begin{overpic}[scale=0.35,tics=20]{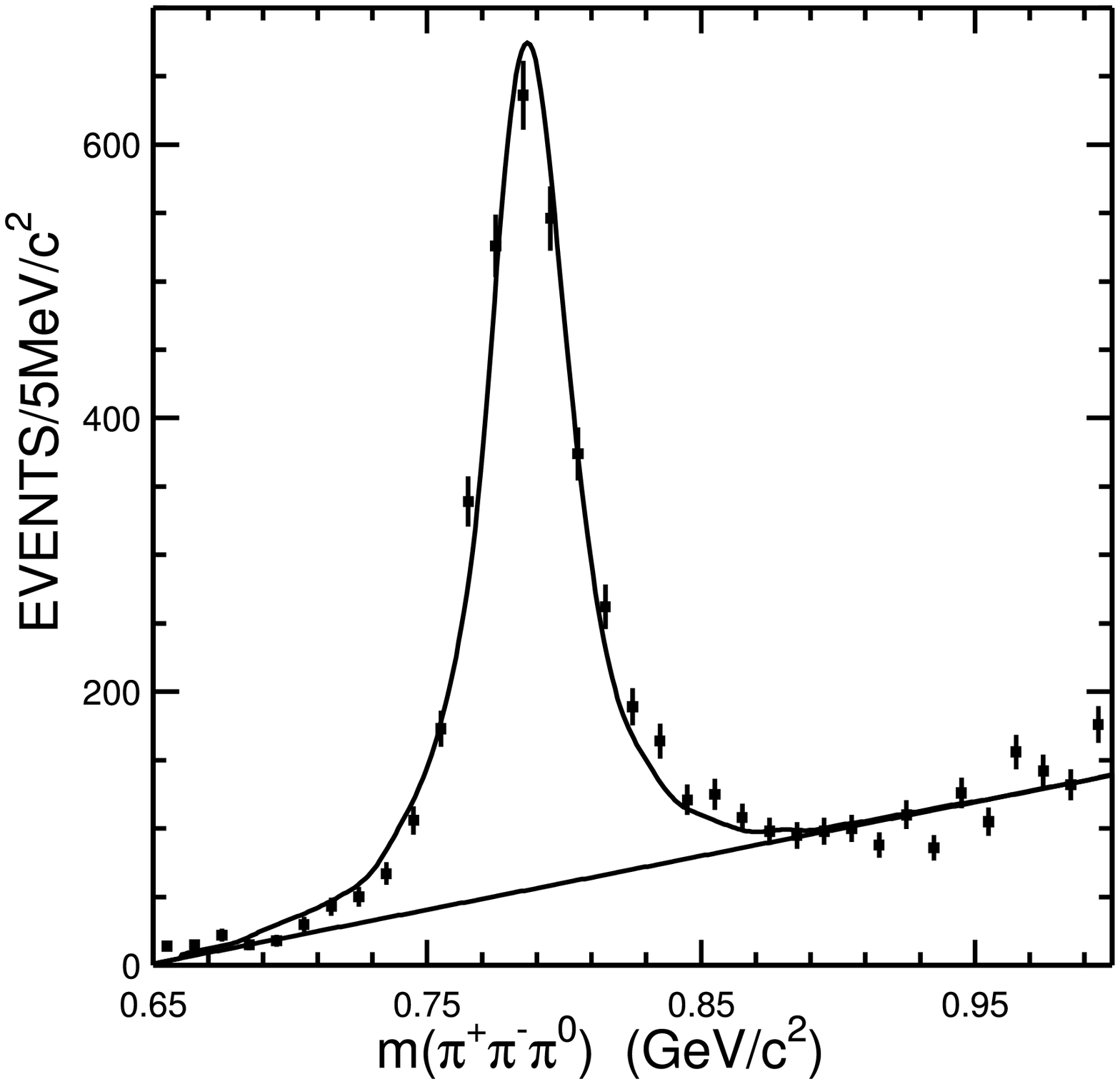}
\put(55,56){
\includegraphics[scale=0.15]{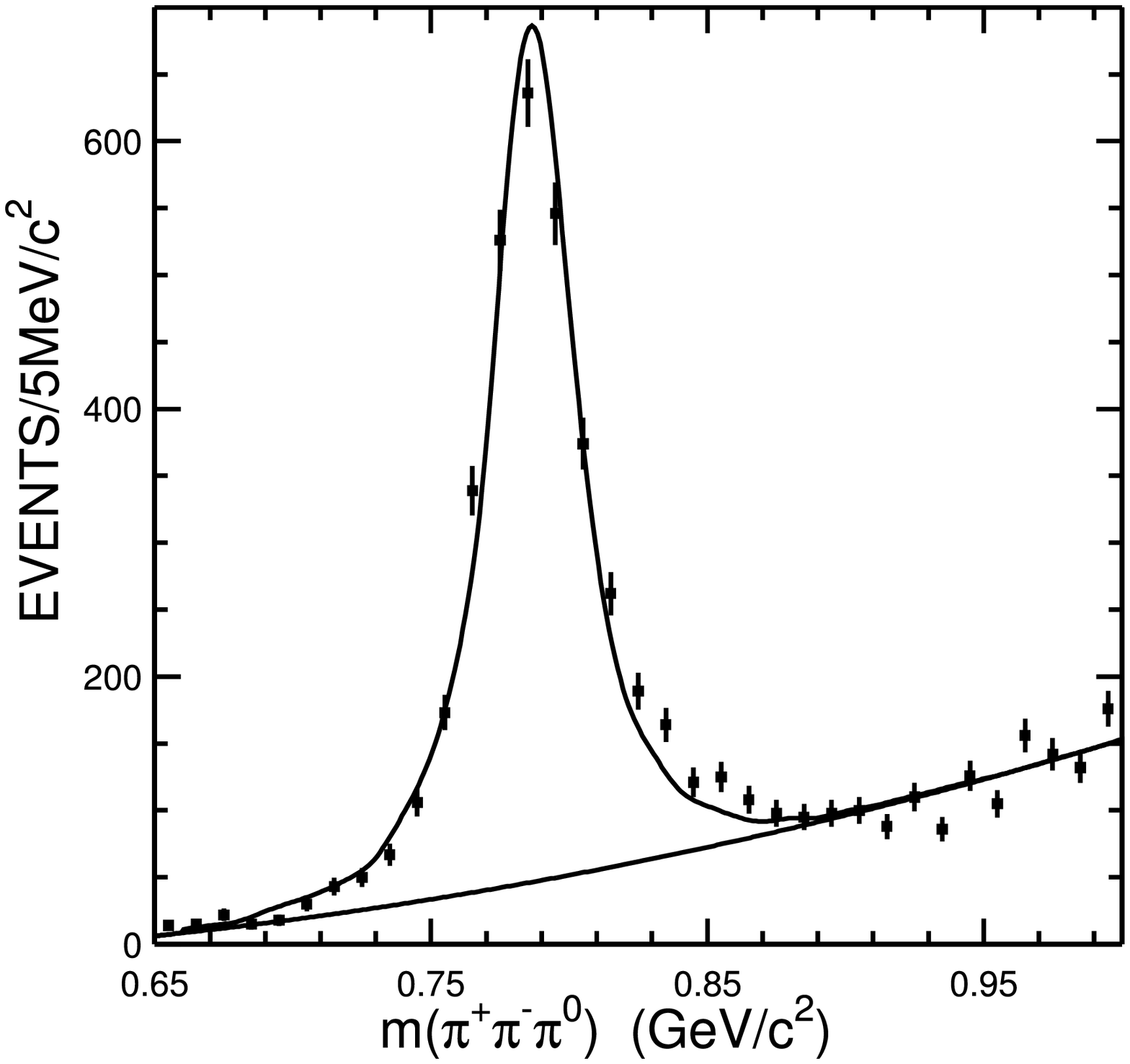}}
\end{overpic}
\caption{The $m_{\pip\pin\pi^0}$ invariant mass distribution for
$\jpsi\ar\omega\pio$ candidate events. The curves are the results of the
fit described in the text. The inset is the fit using a
different background shape (second order polynomial).}
\label{omegapi0}
\end{figure}

The $m_{\pip\pin\pio}$ distribution for events with the recoil $\g\g$
invariant mass ($\g_3\g_4$) being in the $\pio$ mass region,
$|m_{\g_3\g_4}-0.135|<0.04$ GeV/c$^2$, is shown as crosses in
Fig.~\ref{omegapi0}.  The $\omega$ signal, clearly seen in
Fig.~\ref{omegapi0}, is fitted to obtain the branching fraction of
$\jpsi\ar\omega\pio$.  Backgrounds for $\jpsi\ar\omega\pio$ which
contribute to the peak in the $\omega$ signal region mainly come from
non-$\pi^0$ events and events from
$\jpsi\ar\omega\eta(\eta\ar\pio\pio\pio)$ and $\omega\pio\pio$ that
survive selection criteria.  Non-$\pi^0$ events can be measured using
$\pi^0$ sideband events (0.25 $<m_{\g_3\g_4}<$ 0.40 GeV/c$^2$). These
backgrounds will be subtracted after the fit.

%Fig.~\ref{chisq1} shows the 4C $\chi^2$ distributions for
%$\jpsi\ar\omega\pio$, where the crosses are data and the histogram represents 
%the sum of MC simulation of the signal channel $\jpsi\ar\omega\pi^0$ and 
%the normalized backgrounds from $\pi^0$ sidebands, as well as from
%$\jpsi\ar\omega\eta(\eta\ar\pio\pio\pio)$ and $\jpsi\ar\omega\pio\pio$.
%They agree with each other reasonably. 

%\begin{figure}[htbp]
%\centerline{\hbox{\psfig{file=chisq1.eps,width=7cm,height=7cm}}
%}
%\caption{The distribution of 4C $\chi^2$. Where the crosses
%are data, the full histogram is the sum of MC simulation of
%$\jpsi\ar\omega\pio$ and backgrounds from $\pio$
%sidebands, as well as $\jpsi\ar\omega\eta(\eta\ar\pio\pio\pio)$ and
%$\jpsi\ar\omega\pio\pio$.}
%\label{chisq1}
%\end{figure}

\begin{figure}[htbp]
\centerline{\hbox{\psfig{file=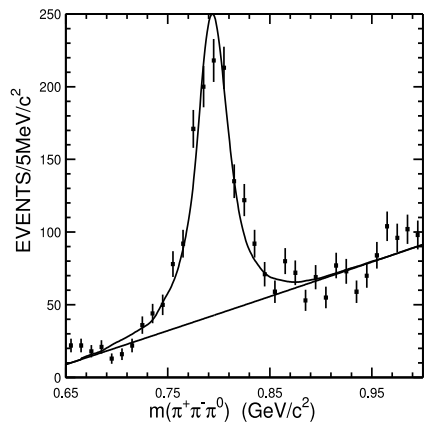,width=5.5cm,height=5.5cm}}
}
\caption{The $m_{\pip\pin\pio}$ distribution for events recoiling
  against the $\pio$ sideband region (0.25 $<m_{\g_3\g_4}<$ 0.40
  GeV/c$^2$).}
\label{pi0sideband}
\end{figure}

A fit to the $m_{\pip\pin\pio}$ distribution is performed using the
expected $\omega$ shape obtained from MC simulation and a first order
polynomial background, shown as the curve in Fig.~\ref{omegapi0}, and
$2595\pm59$ $\omega$ events are obtained. The inset in
Fig.~\ref{omegapi0} shows the fit with a second order polynomial
background.  The $m_{\pip\pin\pio}$ distribution for events which
recoil against the $\pio$ sideband region (0.25 $<m_{\g_3\g_4}<$ 0.40
GeV/c$^2$), shown in Fig.~\ref{pi0sideband}, is fitted to determine
the non-$\pi^0$ background; after normalization, $242\pm10$
non-$\omega$ background events are obtained and subtracted.  We also
subtract $142\pm 18$ background events from
$\jpsi\ar\omega\eta(\eta\ar\pio\pio\pio)$ and $121\pm25$ from
$\jpsi\ar\omega\pio\pio$, which are estimated from Monte Carlo
simulation.

Since $J/\psi \to \omega \pi^0$ is an isospin violation
process, the continuum contribution might be sizable in this channel,
while it can be neglected in $J/\psi \to \omega \eta$ and $\omega \eta'$
decays. 
To determine background contamination from continuum production, the
$L=2347.3$ nb$^{-1}$ data sample taken at $\sqrt{s} =$ 3.07 GeV is
analyzed using the same event selection, and after normalization
$53\pm22$ continuum background events are estimated.  The continuum
contribution is subtracted without considering possible interference.
Backgrounds from other channels are negligible. The detection
efficiency for the signal is $7.55\%$, which is determined by MC
simulation, and the branching ratio for this channel is

\begin{center}
$B(\jpsi\ar\omega\pio)=(5.38\pm0.12)\times 10^{-4}.$
\end{center}
Here, the error is statistical only. 

This decay is an isospin-violating, electromagnetic process,
and we calculate the electromagnetic form factor using the above
branching ratio, according to the formula
\begin{center}
${\frac{|f(m^2_{\jpsi})|^2}{|f(0)|^2}}={\frac{\alpha}{3}}\cdot[\frac{p_\g}{p_\omega}]^3
\cdot\frac{m_{\jpsi}\Gamma(\jpsi\ar\omega\pio)}{\Gamma(\jpsi\ar\g\pio)\cdot\Gamma(\jpsi\ar
\mu^+\mu^-)}$.
\end{center}
It gives 
${|f(m^2_{\jpsi})|}/{|f(0)|}=0.0411\pm0.0009$, which is
consistent with that of MarkIII~\cite{mark3} but is three times
smaller than that of $\jpsi\ar p \bar{p}$~\cite{factor}.

\subsubsection{$\jpsi\ar\omega\eta$}

\begin{figure}[htbp]
\centerline{\hbox{\psfig{file=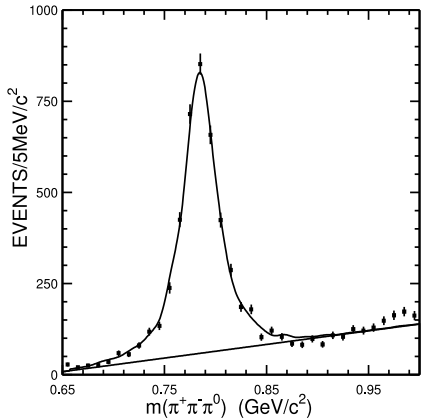,width=5.5cm,height=5.5cm}}
}
\caption{The  $m_{\pip\pin\pi^0}$ distribution for $\jpsi\ar\omega\eta$
  candidate events.
The curves are the results of the
fit described in the text.}
\label{omegaeta1fit}
\end{figure}

The $m_{\pip\pin\pi^0}$ distribution for events where the $\g\g$
invariant mass is required to be in the $\eta$ mass region,
$|m_{\g_3\g_4}-0.547|<0.04$ GeV/c$^2$, is shown in Fig.
\ref{omegaeta1fit}.  A clear $\omega$ signal can be seen.  The main
background events for $\jpsi\ar\omega\eta$ come from non-$\eta$ events
and the events from $\jpsi\ar\omega\eta$, $\eta\ar\pio\pio\pio$ and
$\jpsi\ar\omega\pio\pio$.  Fitting the $m_{\pip\pin\pi^0}$
distribution in Fig.~\ref{omegaeta1fit} with the $\omega$ shape from
Monte Carlo simulation plus a first order polynomial background gives
$3790\pm 72 $ candidate $\omega$ events.  Using the same procedure to
fit the $m_{\pip\pin\pi^0}$ distribution recoiling against the $\eta$
sideband (0.65 $<m_{\g\g}<0.80$ GeV/c$^2$) and normalizing, $188\pm
18$ non-$\eta$ background events are estimated. We also subtract the
$161\pm 17$ and $30\pm 4$ background events from $\jpsi\ar\omega\eta$,
$\eta\ar\pio\pio\pio$ and $\jpsi\ar\omega\pio\pio$, respectively,
which are estimated by MC simulation. The detection efficiency for
$\jpsi\ar\omega\eta$, $\eta\ar\g\g$ determined from MC simulation is
$7.45$\%; thus the $\jpsi\ar\omega\eta$ branching fraction is
\begin{center}
 $B(\jpsi\ar\omega\eta)=(22.86\pm 0.43)\times 10^{-4}$,
\end{center}
where the error is statistical only.

\begin{figure}[htbp]
\centerline{\hbox{\psfig{file=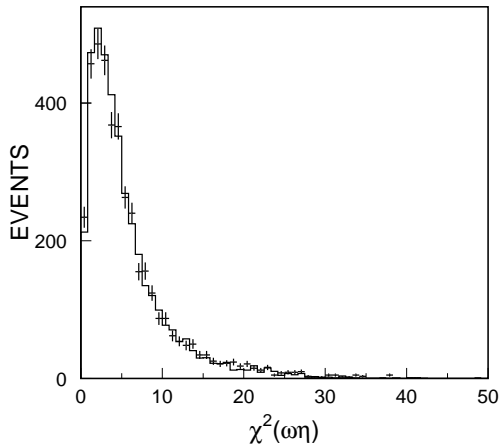,width=7cm,height=7cm}}
}
\caption{The $\chi^2$ distributions for the 4C fit to
the $J/\psi \to \pi^+ \pi^- \gamma\gamma\gamma\gamma$ hypothesis for
$J/\psi \to \omega \eta$ candidate events. The crosses are data; the
full histograms are the sum of MC simulation of $\jpsi\ar\omega\eta$
and non-$\eta$ background determined from $\eta$ sidebands, as well as
from $\jpsi\ar\omega\eta(\eta\ar\pio\pio\pio)$ and
$\jpsi\ar\omega\pio\pio$.}
\label{chisq2}
\end{figure}

Fig.~\ref{chisq2} shows the $\chi^2$ distributions for the 4C fits to
the $J/\psi \to \pi^+ \pi^- \gamma\gamma\gamma\gamma$ hypothesis for
$\jpsi\ar\omega\eta$ candidate events ($|m_{\g_3\g_4}-0.547|<0.04$
GeV/c$^2$ and $|m_{\pip\pin\pio}-0.782|<0.04$ GeV/c$^2$), where the
crosses are data and the histogram is the sum of MC simulation of the
signal channel $\jpsi\ar\omega\eta$ and the backgrounds from
non-$\eta$ events, measured using $\eta$ sidebands, as well as from
$\jpsi\ar\omega\eta(\eta\ar\pio\pio\pio)$ and
$\jpsi\ar\omega\pio\pio$.  They agree with each other reasonably well.

\subsection{\boldmath$\jpsi\ar\omega\g\pip\pin$}
\label{phietag2pi}
For $J/\psi \to \omega \eta$, $\eta \to \gamma \pi^+ \pi^-$, events
with four well-reconstructed charged tracks and at least three
isolated photons are required.  A 4C-fit is performed to the $J/\psi
\to \pi^+ \pi^- \pi^+ \pi^- \gamma \gamma \gamma$ hypothesis, and
$\chi^2 < 20$ is required.  There are 12 possible ways to combine the
charged pions and gammas in forming the $\omega$ and $\eta$ or
$\etap$.

\subsubsection{$\jpsi\ar\omega\eta$}
\begin{figure}[htbp]
\centerline{\hbox{\psfig{file=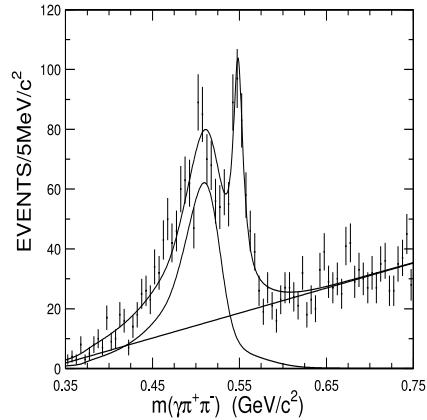,width=5.5cm,height=5.5cm}}}
\caption{Distribution of $m_{\g\pip\pin}$ for $\jpsi\ar\omega\pip\pin\g$
candidate events. The crosses 
are data, and the curves are the results of the fit described in the
text. }
\label{mgpipi}
\end{figure}

Figure~\ref{mgpipi} shows the $\g\pip\pin$ invariant mass recoiling
against the $\omega$ mass region, defined by
$|m_{\pip\pin\pio}-0.782|<0.04$ GeV/c$^2$. A clear $\eta$ signal is
observed.  The enhancement on the left side of the $\eta$ in
Figure~\ref{mgpipi} comes from
$\jpsi\ar\omega\eta~(\eta\ar\pip\pin\pio)$ with one photon missing.
This interpretation as well as the asymmetric shape are confirmed by
MC simulation.

The $\g\pip\pin$ mass distribution is then fitted by this enhancement
with the shape determined from MC simulation, a Breit-Wigner to describe
the $\eta$ signal, and a first order polynomial background.  The fit,
shown in Figure~\ref{mgpipi}, yields $284 \pm 24$ $\eta$ candidate
events.  The contribution of the enhancement is consistent with the
branching ratio for $\jpsi\ar\omega\eta~(\eta\ar\pip\pin\pio)$.
Fitting the $\g\pip\pin$ mass distribution of events recoiling against
the $\omega$ sideband region (0.85 $<m_{\pip\pin\pio}<1.0$ GeV/c$^2$)
and normalizing,
$17\pm 6$ non-$\omega$ background events are estimated
and are subtracted.  The detection efficiency obtained
from MC simulation is $4.59\%$, and the corresponding branching
fraction is
\begin{center}
$B(\jpsi\ar\omega\eta)=(24.47\pm 2.07)\times 10^{-4},$
\end{center}
where the error is statistical.

\subsubsection{$\jpsi\ar\omega\etap$}

\begin{figure}[htbp]
\centerline{\hbox{\psfig{file=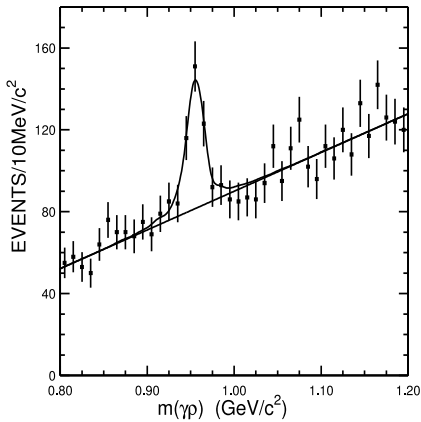,width=5.5cm,
height=5.5cm}}}
\caption{Distribution of $m_{\g\pip\pin}$ for  $\jpsi\ar\omega\rho\g$
  candidate events; the curves are the result of
  the fit described in the text.}
\label{fitomegaetapg2pi}
\end{figure}

After requiring $|m_{\pip\pin\pio}-0.782|<0.04$ GeV/c$^2$ using one
pair of charged pions and the other pair of charged pions to be in the
$\rho$ mass region (0.45 $<m_{\pip\pin}<0.92$ GeV/c$^2$), the
distribution of $\g\pip\pin$ invariant mass recoiling against the
$\omega$ mass region is shown in Figure~\ref{fitomegaetapg2pi}, where
a clear $\etap$ peak is seen.  A fit with the $\etap$ shape determined
from MC simulation and a first order polynomial gives $197\pm 27 $
$\etap$ candidate events.  Fitting the $\g\pip\pin$ invariant mass
distribution of events recoiling from the $\omega$ sideband region
(0.9 $<m_{\pip\pin\pio}<1.0$ GeV/c$^2$) and normalizing, yields
$44\pm11$ non-$\etap$ background events, which are subtracted.  The
detection efficiency obtained from MC simulation is $4.24$\%, and the
branching fraction is
\begin{center}
$B(\jpsi\ar\omega\etap)=(2.41\pm0.33)\times 10^{-4},$
\end{center}
where the error is statistical.

\subsection{\boldmath$\jpsi\ar\omega\pip\pin\g\g$}
For $\eta\ar\pip\pin\pio$ and $\etap\ar\pip\pin\eta$, events with four
well reconstructed charged tracks and at least three isolated photons
are selected. If there are four or more isolated photons, 4C
kinematic fits to the $\pip\pin\pip\pin\g\g\g\g$ hypothesis are made. If
there are only three isolated photons, a 1C kinematic fit with a
missing photon is made, and the fit result is used to determine the momentum
and energy of the missing photon.  Because there
are four pions and four photons, there are 24 possible ways to combine
the charged pions and gammas in forming the $\omega$ and $\eta$ or
$\etap$.

%\begin{figure}[htbp]
%\centerline{\hbox{\psfig{file=dalitz3.eps,width=7cm,height=7cm}}}
%\caption{Scatter plot of $m_{\pip\pin\g_1\g_2}$ versus
%$m_{\pip\pin\g_3\g_4}$
%for $\jpsi\ar\omega\eta$ candidate events.} 
%\label{dalitz3}
%\end{figure}

\subsubsection{$\jpsi\ar\omega\eta$}

\begin{figure}[htbp]
\centerline{
\hbox{\psfig{file=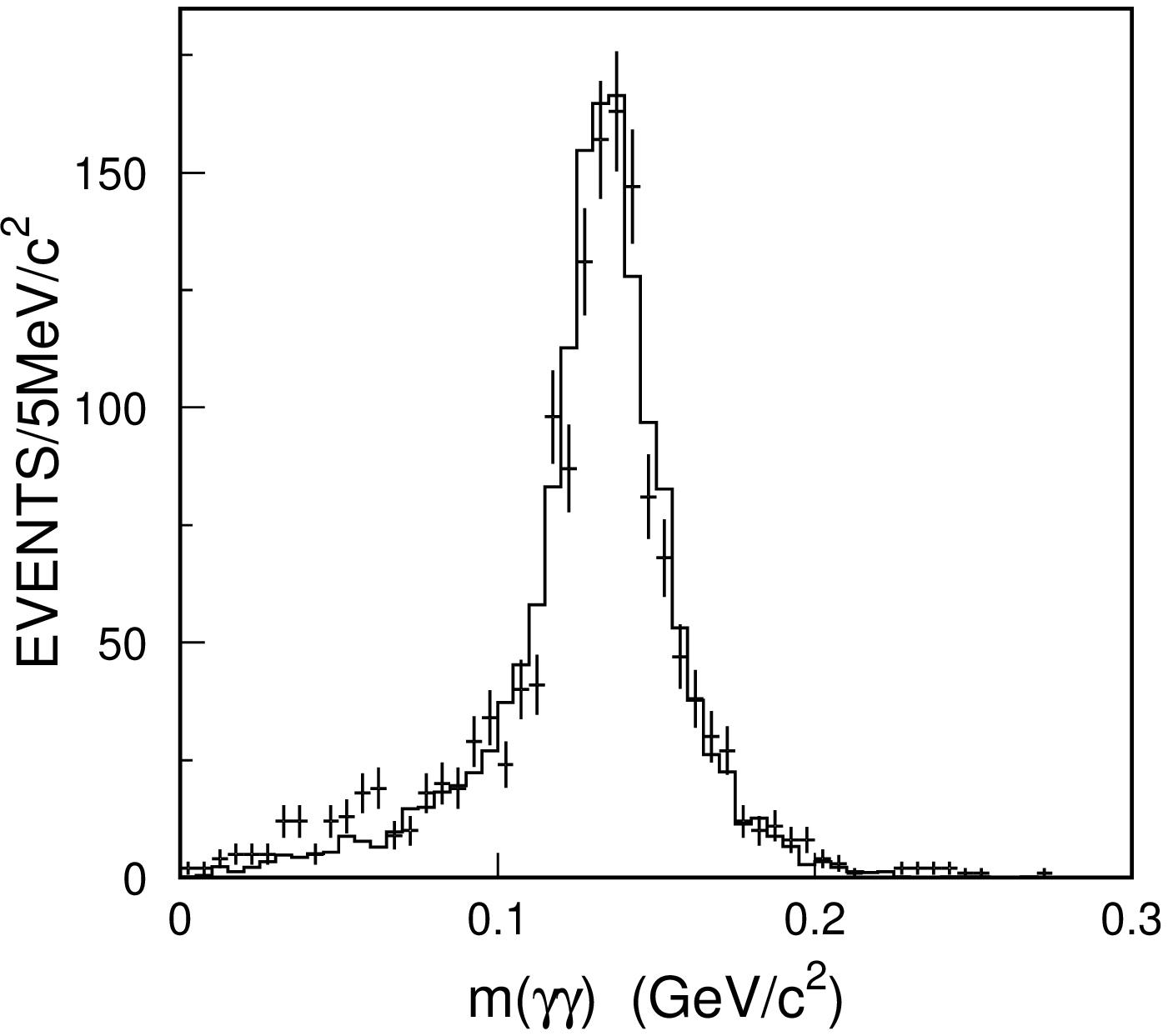,width=5.5cm,height=5.5cm}}}
%\hbox{\psfig{file=g3g4.eps,width=4cm,height=4cm}}}
\caption{The $m_{\g_1\g_2}$ invariant mass distribution for events
  satisfying $|m_{\pi^+\pi^-\gamma_1\gamma_2} - m_{\omega}|
< 0.04$ GeV/c$^2$ and $|m_{\pi^+ \pi^- \gamma_3\gamma_4} - m_{\eta}| <
0.04$ GeV/c$^2$ .
The crosses are data and the histogram Monte
Carlo simulation.}
\label{gamma1}
\end{figure}

\begin{figure}[htbp]
\centerline{\hbox{\psfig{file=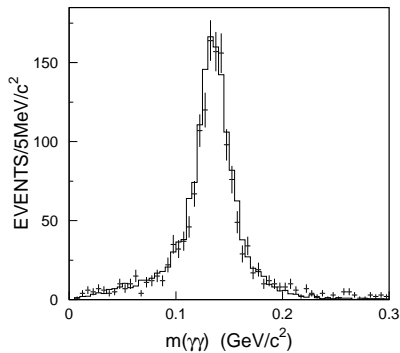,width=5.5cm,height=5.5cm}}}
\caption{The  $m_{\g_3\g_4}$ invariant mass distribution for events
  satisfying $|m_{\pi^+\pi^-\gamma_1\gamma_2} - m_{\omega}|
< 0.04$ GeV/c$^2$ and $|m_{\pi^+ \pi^- \gamma_3\gamma_4} - m_{\eta}| <
0.04$ GeV/c$^2$. The crosses are
data and the histogram  Monte Carlo simulation.}
\label{gamma2}
\end{figure}

Figures \ref{gamma1} and \ref{gamma2} show the $m_{\g_1\g_2}$ and
$m_{\g_3\g_4}$ distributions after the above selection and the
additional requirements $|m_{\pi^+\pi^-\gamma_1\gamma_2} - m_{\omega}|
< 0.04$ GeV/c$^2$ and $|m_{\pi^+ \pi^- \gamma_3\gamma_4} - m_{\eta}| <
0.04$ GeV/c$^2$.  Clear $\pi^0$ signals are observed, and the data and
MC agree well.

\begin{figure}[htbp]
\centerline{\hbox{\psfig{file=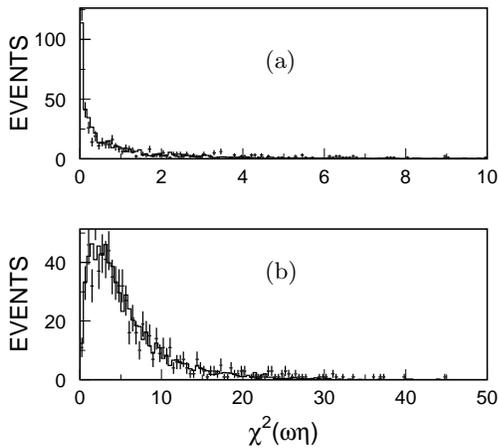,width=7cm,height=7cm}\put(-100,150){(a)}\put(-100,70){(b)}}}
\caption{The $\chi^2$ distributions for the (a) 1C and (b) 4C fits to
the $J/\psi \to \pi^+ \pi^- \pi^+ \pi^- \gamma\gamma\gamma\gamma$
hypothesis for $J/\psi \to \omega \eta, \eta \to \pi^+ \pi^- \pi^0$
candidate events. The crosses are data, and the full histogram is the
sum of the MC simulation of the signal channel and non-$\eta$
background.}
\label{chisq3}
\end{figure}

\begin{figure}[htbp]
\centerline{\hbox{\psfig{file=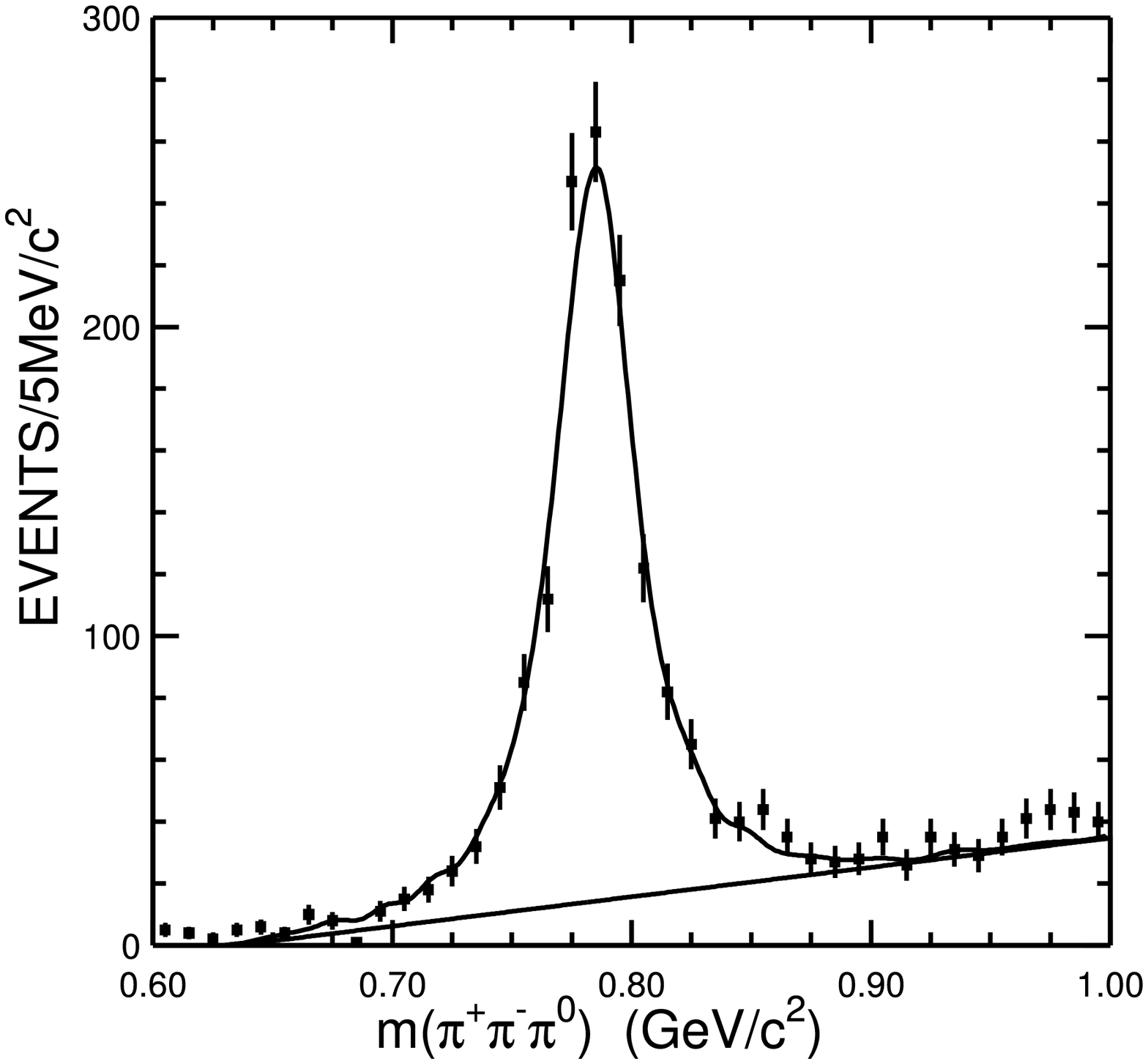,width=5.5cm,height=5.5cm}}
}
\caption{The $m_{\pip\pin\pi^0}$ distribution  for
  $\jpsi\ar\omega \eta$, $\eta \to \pip\pin\pi^0$ candidate events.
The curves are the results of the
fit described in the text.}
\label{weta3}
\end{figure}

The 4C and 1C $\chi^2$ distributions for the fit to
the 
$J/\psi \to \pi^+ \pi^-  \pi^+ \pi^- \gamma\gamma\gamma\gamma$
hypothesis with the requirements that
$m_{\g_1\g_2}$ and $m_{\g_3\g_4}$ are consistent with the mass of the
$\pio$, ($|m_{\gamma \gamma}-0.135|< 0.04$ GeV/c$^2$), and
$m_{\pi^+\pi^-\pi^0}$ is in the $\eta$ mass region ($|m_{\pi^+ \pi^-
\pi^0}-0.547|< 0.04$ GeV/c$^2$), are shown in Fig. \ref{chisq3}.
The ratio of the numbers of events in the two plots for data is
consistent with that from MC simulation.
After the requirements $\chi^2 <20$ for the 4C case and $\chi^2 < 5$
for the 1C case, the $m_{\pip\pin\pio}$ invariant mass spectrum
recoiling against the $\eta$ mass region ($|m_{\pi^+ \pi^-
\pi^0}-0.547|< 0.04$ GeV/c$^2$), shown in Figure~\ref{weta3}, is
obtained.  The $\omega$ shape obtained from MC simulation plus a second
order polynomial are used to fit the $m_{\pip\pin\pio}$ mass
distribution. A total of $1249\pm$43 $\omega$ candidate events is
obtained in the fit. The background determined using the $\eta$
sideband region (0.65 $<m_{\pip\pin\pio}<$ 0.80 GeV/c$^2$) is only
$0\pm2$ events and thus can be ignored.  Using the detection
efficiency of $4.45$\%, the branching fraction of $\jpsi\ar\omega\eta$
is

\begin{center}
$B(\jpsi\ar\omega\eta)=(24.74\pm0.84)\times 10^{-4}$,
\end{center}
where the error is statistical.

\subsubsection{$\jpsi\ar\omega\etap$}

\begin{figure}[htbp]
\centerline{
\hbox{\psfig{file=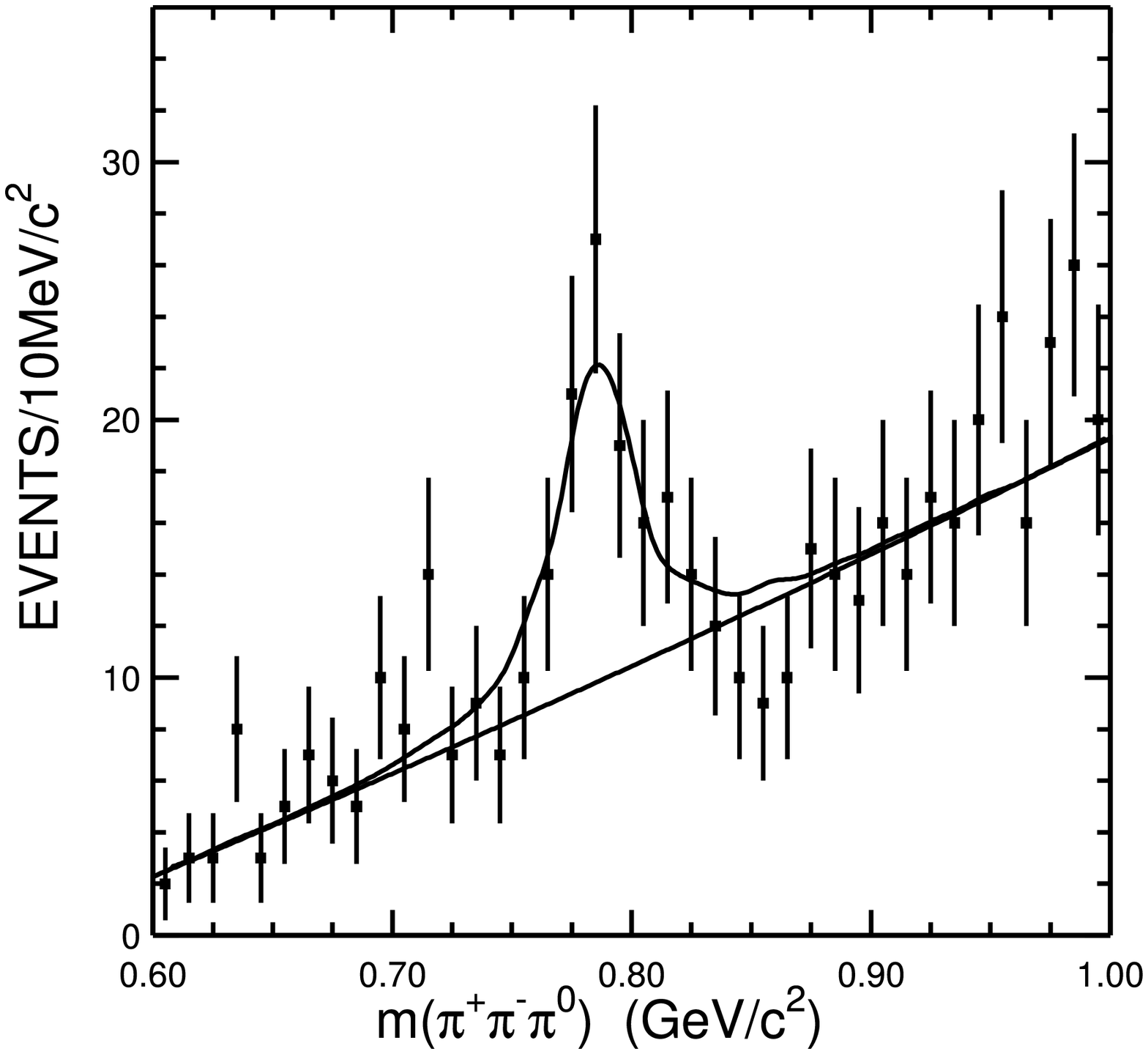,width=5.5cm,height=5.5cm}}}
\caption{The $m_{\pip\pin\pi^0}$ distribution for
  $\jpsi\ar\omega \eta'$, $\eta' \to \pip\pin\eta$ candidate events.
The curves are the result of
the fit described in the text.}
\label{fitomegaetap2pieta}
\end{figure}

After requiring two photons in the $\pi^0$ mass region and other two photons
in the $\eta$ mass region, {\it i.e.}, $|m_{\gamma_1\gamma_2}-0.135| <0.04$
GeV/c$^2$ and $|m_{\gamma_3\gamma_4}-0.547| <0.04$ GeV/c$^2$, as well as
$\chi^2<15$ (5) for the 4C (1C) kinematic fit, the $\pip\pin\pi^0$ mass recoiling against the
$\etap$ mass region ($|m_{\pip\pin\eta}-0.958|<0.04$
GeV/c$^2$), shows a clear $\omega$ peak, as seen in 
Figure~\ref{fitomegaetap2pieta}. The fit of $m_{\pip\pin\pi^0}$ yields
$65\pm 15$ $\omega\etap$ events. No events 
($0\pm1$) are observed in the distribution of events recoiling against the
$\etap$ sidebands (1.10 GeV/c$^2<m_{\pip\pin\eta}<1.15$ GeV/c$^2$) and are
therefore ignored. Other backgrounds are also negligible.
With the detection efficiency for this channel being $3.56$\%, we obtain
\begin{center}
$B(\jpsi\ar\omega\etap)=(2.06 \pm0.48)\times 10^{-4},$
\end{center}
where the error is statistical.

\subsection{Systematic Errors} \label{J-sys}
In this analysis, the systematic errors on the branching fractions mainly come 
from the following sources:
%\begin{itemize}

\subsubsection{MDC tracking and kinematic fit}

In order to study the systematic errors from the MDC tracking and
kinematic fit, clean samples, such as $J/\psi \to \rho \pi$, $\Lambda
\bar \Lambda$, $p \bar p$, $K^* K$, and $\psi(2S) \to \pi \pi J/\psi
(J/\psi \to \mu^+ \mu^-)$, are chosen, and many distributions from
data, including the wire efficiency and resolution of charged tracks,
are compared with those from MC simulations, using two different
treatments of the wire resolution simulation. It is found that in most
cases, the data distributions lie between the two MC
simulations with the different treatments of the wire resolution.
The simulation which agrees better with data is taken as the
official MC simulation, and the difference between the two simulations
is taken as the systematic error for the tracking.

\subsubsection{Particle Identification}

In Refs.~\cite{rhopi2} and \cite{pid}, the particle identification
efficiency of pions is analyzed in detail. Here, only one charged
track is required to be identified as a pion, and
the systematic error from particle identification is less than 1\% and is
negligible.

\subsubsection{Photon detection efficiency}

The photon detection efficiency is studied using $\jpsi\ar\rho^0\pio$
in Ref.~\cite{rhopi2}.  The results indicate that the systematic error
is less than 2\% for each photon. There are slight differences in the
$\pi^0$, $\eta$, and $\omega$ mass resolutions between MC and data.  The
effect of these differences on the branching ratios are very small and
are ignored.
                                                              
\subsubsection{Uncertainty of background}

The background uncertainties come from the uncertainties
associated with the estimation of the sideband backgrounds, the continuum
events, and the events from other background channels, as well as the 
uncertainties of the background shape, different fit ranges, {\it
  etc}.  Therefore, the statistical errors in the estimated background
events, the largest difference in changing the background shape and
the difference of changing the fit range are taken as the systematic
errors from the background uncertainty.

\subsubsection{Intermediate decay branching fractions}

The branching fractions of $\omega\ar\pip\pin\pio$ and the
pseudoscalar decays are taken from the PDG~\cite{pdg2004}. The errors
of these branching fractions are systematic errors in our measurements
and are listed in Table \ref{toterr}.

%\vspace{0.5cm}

The above systematic errors together with the error due to the
uncertainty in the number of $\jpsi$ events are all listed in Table
\ref{toterr}. The total systematic error is determined by adding all
terms in quadrature.

\vspace{0.5cm}
\section{Results}

Table \ref{brpv} lists the branching fractions of $\jpsi$ decaying
into $\omega\pio$, $\omega\eta$, and $\omega\etap$.  The average value is the 
weighted mean of the results from the different decay modes after taking out 
the common systematic errors, and the PDG value is the world
average taken from Ref.~\cite{pdg2004}. The results are higher than
those in the PDG as are our measurements of
$\jpsi\ar\pip\pin\pio$~\cite{rhopi2} and $\jpsi\ar\phi
P(\pi^0,\eta,\etap)$~\cite{phip}. 
It emphasizes the importance of measuring the other decay modes of $\jpsi\ar VP$,
such as $\jpsi\ar\rho\eta$, $\rho\etap$, and $K^*K$ based on the BESII $5.8 \times
10^7 \jpsi$ events.

\begin{table*}[h]
\caption{ Systematic error sources and their contributions. }
\begin{center}
\begin{tabular}{ l|c|c|c|c|c|c}
\hline
\hline
$\jpsi\ar$ &$\omega\pio$ &\multicolumn{3}{|c|}{$\omega\eta$} &
\multicolumn{2}{|c}{$\omega\etap$} \\
\hline
Final state &$\pip\pin\g\g\g\g$ &$\pip\pin\g\g\g\g$
&$\pip\pin\pip\pin\g\g\g$ &$\pip\pin\pip\pin\g\g\g\g$&
$\pip\pin\pip\pin\g\g\g$ &$\pip\pin\pip\pin\g\g\g\g$\\
\hline
Error sources & \multicolumn{6}{|c}{Relative Error (\%)}\\
\hline
%Continue Energy Region & $\sim$1 & $\sim$1  & $\sim$1 & $\sim$1 & $\sim$1
%&$\sim$1 \\
                                                                                                                       
Wire resolution &6.9 & 9.1 & 11.6  & 11.3 & 13.3 &10.3\\
%MDC Tracking &$\sim$4 & $\sim$4 & $\sim$8  & $\sim$8 & $\sim$8 & $\sim$8\\
                                                                                                                       
Particle ID & $<$1& $<$1  & $<$1 & $<$1 & $<$1 & $<$1\\
Photon efficiency & $\sim$8 & $\sim 8 $  & $\sim$ 6 & $\sim 8$ & $\sim 6$
&$\sim 8$\\
%Selection Criteria & 2.44 & 2.44 & 2.81 &1& 2.44&2.93\\
                                                                                                                       
%MC sample &1.20 &1.18 & 1.52  & 2.07 & 1.19 & 1.63\\
                                                                                                                       
%MC Model & 3 & 3 & 3  & 3 & 3 & 3\\

Back. uncertainty &3.3 & 1.0 & 2.42 & 1.0 & 6.6 & 9.7\\
                                                                                                                       
Intermediate decays & 0.79 & 1.05 & 2.48 & 1.95 & 3.48 & 3.48\\
                                                                                                                       
Total $\jpsi$ events& 4.72 & 4.72 & 4.72  & 4.72 & 4.72 & 4.72\\
\hline
Total systematic error & 12.1 & 13.1 & 14.3 & 14.8 & 17.1 & 17.3\\
\hline
\hline
\end{tabular}
\label{toterr}
\end{center}
\end{table*}

\begin{table}[h]
\caption{Branching fractions of $\jpsi\ar\omega\pio$, $\omega\eta$, and
$\omega\etap$.}
\begin{center}
\begin{tabular}{ l |c|c }
\hline
\hline
$\jpsi\ar$          & Final states &Branching Fraction
($\times 10^{-4}$) \\
\hline
$\omega\pio$& $\pip\pin\g\g\g\g$ & 5.38$\pm$0.12$\pm$0.65 \\
              &PDG       & $4.2\pm0.6$ \\
\hline
             &$\pip\pin\g\g\g\g$ &  22.86$\pm$0.43$\pm$2.99 \\

             &$\pip\pin\pip\pin\g\g\g$ & $24.47\pm$2.07$\pm$3.50 \\
$\omega\eta$   & $\pip\pin\pip\pin\g\g\g\g$ & 24.74$\pm$0.85$\pm$3.66  \\
              & Average   &23.52$\pm$2.73 \\
              & PDG       &15.8$\pm$1.6 \\
\hline
              & $\pip\pin\pip\pin\g\g\g$ & 2.41$\pm$0.33$\pm$0.41\\

$\omega\etap$   & $\pip\pin\pip\pin\g\g\g\g$ & 2.06$\pm$0.48$\pm$0.36\\

              & Average              & 2.26$\pm$0.43\\
              & PDG                  & 1.67$\pm$0.25 \\
\hline
\hline
\end{tabular}
\end{center}
\label{brpv}
\end{table}

%\vspace{1.0cm}
\acknowledgments

The BES collaboration thanks the staff of BEPC and computing center 
for their hard
efforts. This work is supported in part by the National Natural
Science Foundation of China under contracts Nos. 10491300,
10225524, 10225525, 10425523, the Chinese Academy of Sciences under
contract No. KJ 95T-03, the 100 Talents Program of CAS under
Contract Nos. U-11, U-24, U-25, and the Knowledge Innovation
Project of CAS under Contract Nos. U-602, U-34 (IHEP), the
National Natural Science Foundation of China under Contract No.
10225522 (Tsinghua University), and the Department of Energy under
Contract No.DE-FG02-04ER41291 (U Hawaii).


\begin{thebibliography}{120}
\bibitem{rhopi} The angular distribution is described by
\begin{center}
${\frac{d^3\sigma}{d\cos\theta_V d\cos\theta_1 d\phi_1}}
={\sin^2\theta_1[1+\cos^2\theta_V+\sin^2\theta_V\cos(2\phi_1)]}$
\end{center}
where $\theta_V$ is the angle between the vector meson and the positron
direction. $\theta_1$ and $\phi_1$ describe the decay products of the
vector
meson in its helicity frame.
\end{thebibliography}
\end{document}